\documentclass[summary]{ursi}
\usepackage{graphicx}
\usepackage{verbatim}
\usepackage{amssymb} 
\usepackage{enumitem}
\usepackage{tabularx}
\usepackage{natbib} 
\usepackage{color, colortbl}
\definecolor{Gray}{gray}{0.9}
\definecolor{pearl}{rgb}{0.94, 0.92, 0.84}
\definecolor{darkgreen}{rgb}{0.0, 0.2, 0.13}
\definecolor{amber}{rgb}{1.0, 0.49, 0.0}
\definecolor{or}{rgb}{0.99, 0.84, 0.69}
\usepackage{amsmath}
\usepackage{graphicx}
\usepackage[colorinlistoftodos]{todonotes}
\usepackage[colorlinks=true, allcolors=blue]{hyperref}
\usepackage{caption}
\usepackage[flushleft]{threeparttable}
\usepackage{siunitx}
\usepackage{float}

\title{Approaches to High Dynamic Range Imaging - Application to the ngVLA}

\author{T. K. Sridharan\affref{ref1}, Sanjay Bhatnagar\affref{ref2}, Preshanth Jagannathan\affref{ref2}  and Kumar Golap\affref{ref2}}

\affiliation{%
  \aff{ref1}{National Radio Astronomy Observatory, VA, USA, tksridha@nrao.edu} \aff{ref2}{National Radio Astronomy Observatory, NM, USA, sbhatnag@nrao.edu, pjaganna@nrao.edu, kgolap@nrao.edu}
}

\begin{document}

\maketitle

\begin{abstract}
  The ngVLA is a new interferometric radio astronomy facility with transformative capabilities, being developed by the National Radio Astronomy Observatory. It combines two orders of magnitude in frequency coverage, over 1.2 - 116 GHz, with unprecedented sensitivity, spatial resolution and spatial frequency coverage, opening up new discovery space, impacting nearly every area of astrophysics. The high sensitivity that enables the path breaking science goals, which in turn lead to stringent instrument requirements, also open up new approaches to meeting them, previously only possible in limited contexts. Chief among the requirements are the image dynamic range specifications of 45 dB and 35 dB at 8 GHz and 27 GHz in single pointing and mosaiced observations. As the baseline calibration strategy to meet these requirements, we leverage the high ngVLA sensitivity through routine use of self-calibration on short time scales to counter atmospheric delay fluctuations and pointing self-calibration to correct for pointing errors. A key benefit of leveraging self-calibration techniques, where possible, is the a reduction in system complexity of a range of subsystems, which in turn improves system reliability. Self-calibration also promises the possibility of attaining thermal noise limited dynamic range performance in some cases. This presentation provides the bases for these approaches, illustrating them to make the case for their application to the ngVLA in parallel. 
  
\end{abstract}

\section{Background and Motivation} 

The science goals of modern interferometers such as the ngVLA require the ability to detect weak emission in the presence of strong peak emission at levels that demand very high dynamic ranges in the constructed images. The emission occurring within an observing field can be exploited to meet the specified dynamic range requirements through self-calibration. 

We first recognize the broad nature of the problem - requiring a certain dynamic range, $DR$, at a targeted science noise level  $\sigma_{science}$ presupposes the presence of bright emission in the field at a corresponding level of detected interferometric flux of $\sim DR \times\sigma_{sci}$, by definition. It is such bright emission  that drives the most stringent dynamic range requirements in the first place, e.g. 45 dB at 8 GHz for the ngVLA targeting deep-field continuum studies of MW-like galaxies, in the presence of a brighter background galaxy source in the field. Would that emission allow self-cal? With the problem posed broadly in this way, one can derive a general limit without recourse to background source counts and the attendant Poisson fluctuations of their occurrence, and which depends only on the number of antennas in the array and the science noise level of the observation, independent of the observing band, primary beam size and antenna SEFDs. This implies a limit on the solution interval at which self-cal can be carried out. This limit is derived below. 

The approach is not fundamentally new, known to many in one form or another and is captured in the maxim roughly stated as  ``{\it ...if you are DR limited, you can selfcal and selfcal will help; if you benefit from selfcal you are (were) DR limited; conversely, if there is not enough SNR for selfcal, you have reached the DR limit your data allows...}'' ({\it e.g.} Bhatnagar, private communication). This synthesis imaging lore and its implications are placed on a formal footing here.

\begin{table*}[h!]
\centering
\label{Tab:PhaseFluctuations}
  \begin{threeparttable}
  \caption{Expected SelfCal solution intervals ($N_A=214$; $SNR_{A\_SelfCal\_threshold}=3; SNR_{map} = 31$).}
  \begin{tabular}{|l|l|l|l|l|l|l|l|l|}
    \hline
    Band & Freq. & DR & $t_{sol}/t_{int}$ & $t_{sol1hr}$ &$t_{sol10hr}$  &$t_{sol16hr}$ & $t_{sol100hr}$ & Driving SCI requirement \\
         & (GHz) &(dB)&                  &   (sec)       &   (sec)       &    (sec)      &    (sec)       &   case, integration time \& rms                       \\
    \hline
    1    & 1.2 & 45 & $9.6\times10^{-7}$&   0.003      &    0.03    &    0.06     &  0.35          &\\
    \hline
    \rowcolor{or}
    2    & 8   & 45 & $9.6\times10^{-7}$&    0.003     &    0.03    &  {\bf 0.06} &  0.35          &SCI-0113, 16hr, 0.035$\mu Jy$ \\
    \hline
    3    & 18  & 40 & $9.6\times10^{-6}$&    0.035     &    0.35    &   0.6      &   3.5         &\\ 
    \hline
    \rowcolor{or}
    4    & 27  & 35 & $9.6\times10^{-5}$& {\bf 0.35}    &    3.5     &   6     &   35       &SCI-0113, 1hr, 0.18$\mu Jy$\\
    \hline
    5    & 50  & 32 & $3.8\times10^{-4}$&   1.4        &   14    &   22      &   140       &\\
    \hline
    6    & 70  & 30 & $9.6\times10^{-4}$&   3.5        &   35     &   56     &   350       &\\
    \hline
    6    & 116 & 28 & $2.4\times10^{-3}$&   8.6       &   86   &   138     &   860      &\\
    \hline
  \end{tabular}
  \begin{tablenotes}
      \tiny
    \item (1) The two frequencies with relevant science requirements derived from key science goals are highlighted, vis., 8GHz: SCI-0113, KSG3-011, 16 hr integration, 0.035$\mu Jy$ and 27 GHz: SCI-0113, KSG3-008, 1 hr integration 0.18 $\mu Jy$.
    \item (2)The applicable $t_{sol}$ limit for these cases are shown in bold.
    \end{tablenotes}
  \end{threeparttable}
\end{table*}

\section{Solution Interval Limit} 

\noindent With $DR, N_A,SNR_{A\_scal\_threshold},t_{sol}$  and $t_{int}$ being  the dynamic range demanded, number of antennas, required SNR per antenna including all baselines to it to allow self-cal, self-cal solution interval lower limit (for a given $DR, SNR_{A\_scal\_threshold}$ and $N_A$) and  integration time to reach a science rms of $\sigma_{science}$, needing to fulfill a demanded dynamic range $DR$ implies a peak flux in the field of 
\begin{equation}
S_p = DR.\sigma_{sci} = DR.SEFD.\sqrt{2}/\sqrt{N_A(N_A-1)\Delta\nu t_{int}}
\end{equation}
Similarly, to achieve a SNR of $SNR_{A\_selfcal\_threshold}$ for one antenna for self-cal to be feasible on a solution interval time scale $t_{sol}$, using all baselines to it, the peak flux needed is
\begin{equation}
S_p =  SNR_{A\_scal\_threshold} \times SEFD/\sqrt{(N_A-1)\Delta\nu t_{sol}}
\end{equation}
\noindent combining (1) and (2) \\
\begin{equation}
t_{sol}/t_{int} = N_A \times (SNR_{A\_scal\_threshold}/DR)^2/2
\end{equation}
\noindent or equivalently,
\begin{equation}
t_{sol} = t_{int}\times N_A \times (SNR_{A\_scal\_threshold}/DR)^2/2
\end{equation}
\noindent $t_{sol}$ is the lower limit to the solution interval above which a SNR of $SNR_{A\_selfcal\_threshold}$ for self-cal to be feasible, is achieved.

While the discussion above is readily appreciated for point sources and peak fluxes, the arguments are equally applicable, and agnostic to, other source structures present. What matters for the signal-to-noise ratio is the interferometric flux actually detected, irrespective of the structures from which it arises. A measure of the detected flux in a general field is the quadrature sum of the emission - as adopted in the ngVLA requirements definition.


Table 1 lists $t_{sol}/t_{int}$  and $t_{sol}$ for $SNR_{A\_selfcal\_threshold} = 3$, for different frequencies for which dynamic range requirements are specified in the ngVLA system requirements and for different values of integration times. The choice of $SNR_{A\_selfcal\_threshold} = 3$ is equivalent to 
full map SNRs of 22 and 31 for 107 and 214 antennas for an integration time of $t_{sol}$. The listed science integration times (or equivalently, science noise levels) of 1, 10 and 100 hrs, span the range of integration times in the relevant science goals which drive the DR specifications. 

 It is formally incomplete to specify a dynamic range without a targeted science noise level (or integration time), as often done in instrument science requirements  – it matters where the specified dynamic range is placed in the range of observed flux levels. This has strong implications for applicable calibration strategies, in particular, if self-cal would be feasible or not. 

As can be seen from the table, as the required dynamic range increases, the feasible solution interval decreases. With a higher demanded dynamic range, the peak emission that demands this dynamic range is higher, allowing a smaller solution interval.  For the two frequencies, 8 \& 27 GHz with the highest DR requirements, the minimum solution interval limits indicated are $<1$s, above which $SNR_{A\_selfcal\_threshold} =$ 3 is achieved, for the noise levels (or equivalently, the integration times) called for by the respective science. In other words, there is sufficient SNR to deliver self-cal on the time scales as short as indicated. This is made possible by the high sensitivity of the ngVLA arising from its large collecting area. Thus, very fine-grained tracking of antenna gains is possible through self-cal. Based on past experience, self-cal is the only time-tested path to achieve high dynamic range imaging 
and should be a key calibration strategy for the ngVLA.  The numbers derived show that this is also feasible: for the time scales implied – as short as 0.06 s and 0.35 s at 8 \& 27 GHz – experience indicates a comfortable situation for attaining high dynamic range imaging. While the most sensitive observations will use the most number of antennas, with only 107 antennas considered, the solution interval time scales are still very short, at 0.12 s and 0.7 s at 8 \& 27 GHz. To be more definitive, these solution intervals should be compared with time scales on which gain fluctuations occur. Instrumental fluctuations are clearly on much longer time scales, leaving only fluctuations of ionospheric and tropospheric origins to consider, which we now turn to. 

\section{Atmospheric Phase Fluctuations}

Our approach is to estimate the residual phase fluctuations expected after calibration on various time scales suggested in Table 1 and assess if they would be sufficiently small. While more details can be found in Carilli \& Holdaway (1997), Carilli et al (1999), Butler \& Desai (1999) and Sridharan \& Bhatnagar (2023), briefly, for fast switching calibration with a calibrator $\theta$ radian away, a cycling time of $t_{cyc}$ and winds aloft of $V_a$, the phase fluctuations are {\it stopped} at an effective baseline scale $b_{eff}$ given by
\begin{equation}
b_{eff} = V_a \times t_{cyc} / 2 + \theta \times h
\end{equation}
\noindent where $h$ is the height at which the dominant phase fluctuations occur. For self-cal, the second term drops out as $\theta \approx 0$ (calibrator is the target, in the primary beam), leading to
\begin{equation}
b_{eff, selfcal} = V_a \times t_{sol} / 2 
\end{equation}
\noindent For the short solution intervals indicated above (0.06 \& 0.35 s; Table 1), and  a nominal $V_a$ of $\sim 10$ m/s (Carilli et al 1999), $b_{eff}$ is $\sim 1-5$ m for 8 and 27 GHz. In other words, the implied residual phase fluctuations of atmospheric origin after self-calibration would be that of an array of only $\sim$ 5-10 m extent. While the strict applicability of the approach to such small scales may be questioned, it is clear that the expected residual fluctuations will be very small.

It is evident from the above discussion and Table 1 that the selfcal solution intervals achievable are orders of magnitude lower than needed, allowing a large margin for the feasibility for self-cal and that selfcal should be the baseline calibration strategy for bands 1-4 (1.2 to 34 GHz) for the ngVLA.

\section{Partially Resolved Sources}

While the arguments above apply irrespective of source structure, the possibility that the emission in the beam may be resolved out on some baselines leading to lower SNR on those baselines needs to be considered.
There are multiple mitigating factors: \\
\noindent (1) As previously noted, the signal to noise ratio and the dynamic range depend on the actual detected fluxes, a measure of which is the quadrature sum of the emission, including distributed emission.   \\
\noindent (2) The baselines on which flux is (partially) resolved out also contribute proportionally smaller image errors, impacting the DR less, by the same factor. The following points and approximate estimates help build some intuition for the impact. In the extreme case, a baseline with no flux does not contribute an image error irrespective of the gain errors present.  If $\sim$ half the antennas have baselines with zero flux, the DR is lowered only by $\sim$ 1.5 dB ({\it i.e.} $\sqrt{N}$; Sridharan et al 2023), although a different total integration time will be required to reach the targeted sensitivity. In general, if a fraction {\it f} of the antennas has baselines with lower fluxes by a factor {\it r} due to resolved source structure, the resulting DR decreases by: 
\begin{equation}
\Delta DR \approx  ( 1- f(1-r))/\sqrt{1-f(1-r^2)}
\end{equation}
\noindent for natural weighting (more details may be found in Sridharan \& Bhatnagar, 2023). For baselines to half the number of antennas losing half the flux, $f = 0.5$ and $r = 0.5$ and DR decreases by 0.2 dB. \\
\noindent (3) Source counts imply some background point sources in the beam at some level. 

When phase fluctuation data are available from a test interferometer (as the case for ngVLA), a complemetary approach can be taken comparing 
the measured fluctuation time scales with the ones implied in section 2 and Table 1. Using such an approach Sridharan \& Bhatnagar show that even with flux resolved out by a factor of $\sim$20, self-cal remains a feasible strategy. Additionally, as outlined above the impact of such
lost flux on image dynamic range is expected to be minimal.

\section{Achievable Dynamic Range}

The next question is what dynamic range can be achieved through self-calibration. As we show below, by definition, this is self-consistently equal to the required dynamic range, $DR$, which we started with, as long as self-cal is feasible on the required time scales. 

The per antenna phase error on the self-cal solutions for the above context ({\it i.e.} the combination  $ SNR_{A\_selfcal\_threshold},DR, t_{sol},t_{int}$ and $\sigma_{science}$), that can be ideally achieved,  is  $\sim 1/SNR_{A\_selfcal\_threshold}$  radian. 
\begin{equation}
\phi_{\sigma,A,selfcal,t_{sol}} =  1/SNR_{A\_scal\_threshold}
\end{equation}
\noindent using (3), we have
\begin{equation}
\phi_{\sigma,A,scal,t_{sol}} =   (N_A \times t_{int}/t_{sol})^{1/2}/DR
\end{equation}
\noindent This would be the residual phase error on each antenna after applying self-cal solutions. The SNR on a map made with baselines to one antenna with these self-cal corrections is, then,
\begin{equation}
SNR_{map,1Ant} =  1/\phi_{\sigma,A,scal,t_{sol}} =  DR /(N_A \times t_{int}/t_{sol})^{1/2}
\end{equation}
\noindent The SNR for the full map, combining all antennas and for the full integration time would be higher by  $(N_A \times t_{int}/t_{sol})^{1/2}$. Therefore, from (10), we have,
\begin{equation}
SNR_{map,scal,t_{int}} = DR 
\end{equation}
\noindent This SNR is the map dynamic range achieved. Thus,
\begin{equation}
DR_{map,scal,t_{int}} = DR
\end{equation}
In short, we reach a general, self-consistent conclusion that as long as the solution interval time scale on which self-cal is feasible is smaller than the time scales of corrupting residual fluctuations, the required dynamic range is achievable.

\section{Other Prospects}

If self-cal is feasible on smaller time scales, as is the case here, other benefits accrue. The correspondence between time and bandwidth can be exploited to split the spectral coverage into sub-bands – it is the $t_{sol} \times \Delta\nu_{sol}$ product, the number of independent data samples, that matters - to handle changing source structure over the band, which would otherwise impact self-cal. Parameterization of the spectral variation may be a more efficient approach. Given the additional room, other wide bandwidth and direction dependent effects would also become amenable to elimination through self-cal. The fact that these effects are only expected to corrupt the observed visibilities on longer time scales and can therefore tolerate longer solution intervals, further eases the situation. Effectively, the high sensitivity of the ngVLA holds out the possibility of reaching thermal noise limited dynamic range, at least in some bands and in some cases.

 Additionally,  the ability to track gains through self-calibration allows the electronics and receiver systems to be designed to less stringent stability specifications and may even offer the potential to eliminate some subsystems entirely. A case in point is the need for a dedicated water vapor radiometer (WVR) for the ngVLA, as envisaged in the baseline calibration approach for delay corrections in band 2 (8 GHz) to deliver the required 45 dB image dynamic range. In the self-calibration approach, this would no longer be needed. This opens the potential for the rearrangement of the receiver locations, allowing the use of the band 4 science receiver, which covers the 22 GHz water line, to be placed between bands 5 \& 6 and thus provide WVR corrections for observations in bands 4, 5 \& 6 ($\sim$ 20-34, 30-50 \& 70-116 GHz). The prospects for this simplification are currently being explored. Thus, leveraging the ability to self-calibrate as the baseline calibration strategy can offer paths for reducing overall system complexity including possible elimination of some signal chains and subsystems.    
\section{Pointing Self-Calibration}

The image dynamic range achieved in an observation, $DR$, improves with the number of antennas $N$ and is compromised by the pointing errors these antennas suffer. We adopt a conservative $\sqrt{N}$ dependence of $DR$ on $N$, as opposed to the conventional $N$ dependence. In the presence of $N$ independent antenna pointing errors, which lead to $N$ independent antenna based visibility amplitude errors, a $\sqrt{N}$
dependence would be expected (Sridharan et al, 2023). With this scaling, the stringent 45 dB and 35 dB image dynamic range specifications require 
sub-arcsec pointing performance on time scales shorter than reference pointing. Our strategy here, again, is to leverage the high sensitivity of the ngVLA by exploiting the pointing self-calibration technique developed by Bhatnagar \& Cornwell (2017) to reach the required $DR$ levels  without requiring such pointing performance. The feasibility of this approach to correcting for pointing errors is currently being evaluated for the ngVLA (Janannathan et al 2023). Once again, the use of pointing self-calibration simplifies the demands on the pointing control system, lowering its complexity.

\section{Conclusion}

In conclusion, self-calibration is feasible for achieving the high image dynamic ranges required to satisfy ngVLA science goals in bands 1-4 ($\sim$ 1.2 - 34 GHz) and is the preferred baseline strategy. Self-calibration also self-consistently allows the attainment of the dynamic range demanded, provided short enough solution intervals are reached. Similarly, pointing self-calibration is being pursued as the method for minimizing the effects of pointing errors on the images. These strategies require high sensitivity for routine use, and therefore, are applicable to the ngVLA and promise the possibility of reaching thermal noise limited dynamic range performance in some cases. The limits from extension of this approach to higher frequencies remain to be worked out, and may only be feasible in limited cases. A key benefit of leveraging self-calibration techniques, where possible, is the the reduction in system complexity of a range of subsystems, which in turn improves system reliability.


\end{document}